\documentclass{article}
\usepackage[cp1251]{inputenc}
\usepackage[english]{babel}
\usepackage{graphicx}
\usepackage{amsmath}
\usepackage{amssymb}
\usepackage{latexsym}
\usepackage{natbib}
\begin{document}
\begin{flushright}
Journal-Ref: Astronomy Letters, 2018, Vol. 44, No. 2, pp. 119-125
\end{flushright}

\begin{center}
\Large {\bf Simulations of the dynamics of the debris disks
in the systems Kepler-16, Kepler-34, and Kepler-35}\\

\vspace{0.5cm}
\large {\bf T.V.\,Demidova, I.I.\,Shevchenko}

\normalsize

Pulkovo Astronomial Observatory of the Russian Academy of Sciences,\\
Pulkovskoe shosse 65, St. Petersburg, 196140 Russia\\
\end{center}
\begin{center}
e-mail:proxima1@list.ru
\end{center}
\normalsize
\begin{abstract}
The long-term dynamics of planetesimals in debris
discs in models with parameters of binary star systems Kepler-16,
Kepler-34 and Kepler-35 with planets is investigated. Our calculations have shown the formation of a stable
coorbital with the planet ring is possible for Kepler-16 and Kepler-35 systems. In Kepler-34 system, significant
eccentricities of the orbits of the binary and planets can
to prevent the formation of such a structure. Detection
circumbinary annular structures in observations of systems
binary stars can be evidence of the existence of planets,
retaining coorbital rings from dust and planetesimals.

\end{abstract}

\section{Introduction}
Modern scenarios for the formation of planets
predict their formation in a gas-rich protoplanetary
disk \citep[see, e.g.,][]{2012RAA....12.1081Z}. However, the
formation of planets in binary star systems involves
a number of difficulties. The theory and numerical
experiments show that gravitational perturbations
in binary star systems cause the eccentricities of
planetesimals to increase periodically. This increase,
in turn, prevents their mutual accretion, because it
leads to their destruction in high-velocity collisions
\citep{2004ApJ...609.1065M, 2012ApJ...752...71M, 2012ApJ...761L...7M, 2012ApJ...754L..16P}. The scenario for
the formation of a planet far from the central instability
region and its subsequent radial migration
inward, toward the central binary seems probable in
that case. In the systems Kepler-16, Kepler-34, and
Kepler-35, in order to the planet to occupy its current
orbital position in the system, during its migration
it would have to cross several zones of chaos attributable
to the orbital resonances with the central
binary \citep{2013ApJ...769..152P}.

The formation of circumbinary planets is most
likely in the case of close binary systems, while wide
star pairs are suitable for planets orbiting around one
of the binary components \citep{1999AJ....117..621H, 2007A&A...469..755M}. Most of the discovered circumbinary
planets belong to binary stars with a separation
of less than 1 AU \citep{2011Sci...333.1602D, 2012Natur.481..475W,2012Sci...337.1511O}.

With time, the gas component
is depleted and the protoplanetary disk passes to
the planetesimal stage. If protoplanets have already
been formed by this epoch, then their presence can
be detected by the interaction with the planetesimal
disk. Observations have shown that the planetesimal
disks of both single and binary stars have nonsymmetric
and ring-like structures, which may suggest
the presence of unresolved companions in the
disk, for example, giant planets or low-mass stars
\citep{1998ApJ...506L.133G, 2004ASPC..321..305A, 2011ApJ...743L...6T, 2012AJ....144...45K, 2013ApJ...762L..21M,2017A&A...600A..72F}. This idea has stimulated the studies aimed at searching for the connection of such
structures with characteristics of the invisible companion,
primarily its mass and orbital parameters. For
example, \citet{2000ApJ...537L.147O} studied the clumps
in the planetesimal disk of a single star caused by
the influence of a planet; planetary perturbations were
shown to produce an asymmetric resonant dust belt
with one or more clumps, intermittent with cavities.
\citet{2016MNRAS.463L..22D} investigated
a similar problem for the case of a circumbinary planetary
system; evidence that the ring-like structure
coorbital with the planet is more massive and stable
if the central star is a binary rather than a single one
was obtained.

The populations of small bodies coorbital with Solar
System planets are well known. Jupiter's Trojans
are a well-known example of the motion of small
bodies in $1:1$ resonance \citep[see, e.g.][]{1999ssd..book.....M}.  \citet{2002Icar..160..271N} presented their simulations of the dynamics of Saturnian, Uranian,
and Neptunian Trojans on time scales comparable
to the age of the Solar System. Their calculations
showed that the coorbital populations of
asteroids are rapidly scattered in the case of Saturn
and Uranus, but Neptune can retain $50\%$ of
its original population of Trojans. Several Trojans
were discovered near Mars \citep{1990BAAS...22.1357B} and
Neptune \citep{2006Sci...313..511S}, one near the
Earth \citep{2011Natur.475..481C}, and one near Uranus \citep{2013Sci...341..994A}. The existence of a dust
ring coorbital with the Earth was predicted by \citet{1989Natur.337..629J}. Subsequently, the ring was
discovered on the basis of data from the IRAS \citep{1994Natur.369..719D} and COBE \citep{1995Natur.374..521R}
space missions. The discovery of such objects and
structures in the Solar System suggests that they can
also be revealed in other planetary systems.

Since, as has been noted above, the coorbital
rings for circumbinary planets are more stable and
long-lived than those for planets of single stars, the
ring-like structures of dust and small bodies can
be revealed near the orbits of discovered circumbinary
planets. In this paper we consider the systems
Kepler-16, Kepler-34, and Kepler-35. Based on our
numerical experiments, we describe the characteristics
of hypothetical coorbital rings.

Note that the planetesimals in our models gravitate
passively, i.e., their masses, sizes, and selfgravity
are disregarded. The masses for the observed
debris disks vary from $\sim 3$ to $\sim 20$ Earth masses \citep{2002MNRAS.334..589W, 2009ApJ...693..734C}. \citet{2014A&A...561A..43B} and \citet{2014MNRAS.443.2541P} showed
that the influence of a massive planet dominates over
the mutual gravitational interaction (self-gravity) of
planetesimals if the planet's mass exceeds the disk
mass by an order of magnitude or more. In all our
models the ratio of the planet's mass and the disk
mass is $\sim 100$. The simulation data on the dynamics of
planetesimal disks with and without self-gravity were
compared by \citet{2015A&A...582A...5L}. They showed that in
the systems Kepler-16 and Kepler-34 the contribution
of self-gravity affects the distribution of matter in
the disk (and the eccentricities of planetesimals) only
slightly.

Under conditions of flat planetesimal disks of
stars, whose gravity dominates in the disk dynamics,
the standard theory of \citet	{1942psd..book.....C}, as \citet{1988PThPS..96..175H} showed, gives not the characteristic
pair gravitational relaxation time itself but
its lower limit. According to \citet{1988PThPS..96..175H},
the lower limit for the relaxation time is estimated
as $\sim 0.005{v^3}_p/G^2n_p{m^2}_p$, where $G$ is the gravitational constant, $m_p$ is the characteristic mass of a gravitating particle, $n_p$ is the space density of interacting particles, and $v_p$ are their characteristic relative velocities.
The characteristic relative velocities of planetesimals
in circumbinary disks determined by the perturbations
from the central binary are $\sim 1000$ m s$^{-1}$ \citep{2004ApJ...609.1065M, 2013A&A...553A..71M}.
For planetesimals with a diameter of $10$ km and a
density of $2$ g cm$^{-3}$ in a disk with a particle space
density $n_p \sim 10^{-24}$ cm$^{-3}$ we have $\sim 10^7$ yr for the lower limit of the relaxation time in our model, which
is much longer than our integration time intervals
but is comparable to the lifetimes of the disks under
consideration. Thus, self-gravity can be an important
effect, especially regarding the long-term evolution of
coorbital structures. We leave this question for future
further studies. The Toomre parameter \citep{1964ApJ...139.1217T} for the disks under consideration is $Q \gg 1$ due to their
low masses. Therefore, we assume that the disk selfgravity
in our models is negligible.

\citet{2012ApJ...754L..16P} and \citet{2012ApJ...752...71M}
investigated the collisional dynamics of planetesimals
in circumbinary disks. According to their calculations,
the collisions of planetesimals at high velocities
in the system Kepler-16 are intensive in the ring between
$1.75$ and $4$ AU. They obtained similar data for
the systems Kepler-34 and Kepler-35 as well. Near
the planets themselves the collisions of planetesimals
most likely do not lead to significant destructions of
planetesimals. Therefore, in all likelihood, collisions
do not affect the dynamics of the coorbital ring. However,
fine dust can be produced in the planetesimal
disk due to collisions. Its presence will allow various
structures in disks to be revealed through infrared
observations.

\citet{2017MNRAS.465.4735M} considered the influence of
self-gravity on the evolution and structure of gas-rich
circumbinary disks using the early epochs
of evolution of Kepler-16, Kepler-34, and Kepler-35
as an example. According to their results, if the disk
is sufficiently light, then self-gravity does not affect
significantly the structure of the disks in Kepler-16
and Kepler-35 with low binary eccentricities, but the
differences can be significant in the case of Kepler-34.
If the gas disk is sufficiently massive, then the disk
self-gravity can affect strongly the disk structure, in
particular, the size and shape of the inner cavity. The
migration of the forming planet can also be affected,
because the inward-migrating planet stops near the
inner disk boundary \citep[the cavity edge;][]{2008A&A...483..633P}. The observed positions of the planets
and the numerically calculated sizes of the inner
cavity were compared by \citet{2017MNRAS.469.4504M}. They
showed that there is no need to take into account the
disk self-gravity to explain the current position of the
planet in Kepler-16, while for Kepler-34 good agreement
is achieved with self-gravity. No agreement was
achieved in any of the models for Kepler-35.

Note also that the possibility of the planet's migration
in the protoplanetary disk of the binary system
is disregarded in our calculations. In principle, the
radial migration of the planet in the disk can prevent
the formation of a ring-like structure coorbital with
the planet.

The migration of planets in a circumbinary disk
takes place mainly at the stage of their formation
when the disk is still gas-rich. \citet{2007A&A...472..993P}  showed that bodies with a mass up to
20 Earth masses migrate in a gas-dust circumbinary
disk inward (toward the binary). Their motion is
stopped near the inner disk boundary, the boundary of
the central cavity. Such objects can become embryos
for more massive planets, because the disk gas can
be accreted onto them. \citet{2008A&A...483..633P}
showed that if the mass of a planet grows to the mass
of Saturn, then the planet can occupy a stable orbit
around the binary. However, a planet with the mass of
Jupiter will most likely be moved to the disk periphery.
Theoretical estimations and numerical simulations of
the dynamics of planets in the circumbinary disks of
Kepler-16, Kepler-34, and Kepler-35 \citep{2012ApJ...754L..16P, 2012ApJ...752...71M, 2014ApJ...782L..11L, 2015A&A...582A...5L,  2016A&A...590A..62L} have shown that the formation of circumbinary
planets in situ (i.e., near their observed positions) is
complicated. Thus, the migration of planets from outside
the system inward, toward the current positions
is needed. These positions roughly correspond to the
centers of the resonance cells (restricted by the integer
orbital resonances of the planet with the binary)
and, therefore, are stable \citep{2013ApJ...769..152P, 2016AstL...42..474P}.

The migration is relatively insignificant at the
stage of a gas-free planetesimal disk. As \citet{2014MNRAS.443.2541P} showed, the migration velocity
of a planet due to the scattering of planetesimals
is determined mainly by the ratio of the masses of
the planet and the debris disk. There is virtually no
migration if this ratio exceeds $\sim 10$. Therefore, there
is reason to believe that in our models, where this ratio
is $\sim 100$, this effect may also be disregarded.

The possibility of the migration of a planet to an
orbit far from the binary was discussed by \citet{2008A&A...483..633P} and \citet{2017A&A...602A..12R}. Under
such migration the formation of a long-lived coorbital
ring can probably be complicated. On the whole, the
problem of the influence of the migration of planets
on the survival of their coorbital structures requires
additional studies.

\section{The model}
We consider a model of a binary star system with
component masses $M_1$ and $M_2$ with a semimajor axis
$a_b$ and eccentricity $e_b$. The binary is embedded in a planetesimal disk with a radius of $4$ AU. The model
parameters are given in Table~\ref{tab:models}. They correspond
to Kepler-16 (model 2), Kepler-34 (model 4), and
Kepler-35 (model 5) \citep[see][]{2011Sci...333.1602D, 2012Natur.481..475W}. For comparison, we consider a model of
a binary star system with the parameters of Kepler-16
without a planet (model 1) and a system with a planet
around a single star with a mass equal to the sum of
the masses of the Kepler-16 stars (model 3).
\begin{table} [t]
   \vspace{6mm}
    \centering
    \caption{ {\bf The model parameters}}
    \label{tab:models}
    \vspace{5mm}\begin{tabular}{l|c|c|c|c|c} \hline\hline
    Model & 1 & 2 (K-16) & 3 & 4 (K-34) & 5 (K-35)  \\
    \hline
    $M_1$ $[M_\odot]$  & 0.690 & 0.690 & 0.893 & 1.048 & 0.888 \\
    $M_2$ $[M_\odot]$ & 0.203 & 0.203 & - & 1.021 & 0.809  \\
    $a_b$ [AU] & 0.224 & 0.224 & - & 0.229 & 0.176   \\
    $e_b$ & 0.159 & 0.159 & - & 0.521 & 0.142 \\
    $m_p$ $[10^{-4}M_\odot]$ & - & 3 & 3 & 2.1 &  1.2   \\
    $a_p$ [AU] & - & 0.705  & 0.705 & 1.09 &  0.603   \\
    $e_p$ & - & 0.007 & 0.007 & 0.182  &  0.042   \\
    $\gamma$ $[^\circ]$ & - & 69.39 & 69.39 & 176.96 &  24.14   \\
    \hline
    \end{tabular}
\end{table}
We investigate the dynamics of negligible-mass
planetesimals in the gravitational field of a binary star
and a circumbinary planet. The planet has mass $m_p$,
orbital semimajor axis $a_p$, and eccentricity $e_p$. The
initial angle between the directions to the pericenters
of the binary and planet orbits is denoted by $\gamma$. Twenty
thousand planetesimals initially placed in circular orbits
with Keplerian velocities are simulated. The
radial distribution of planetesimals is specified by the
law $a^{-1}$ ($a$ is the orbital radius).

The equations of planetesimal motion in the
barycentric coordinate system are written as
\begin{equation}
   \frac{d\vec{v}}{dt} = \nabla\Phi_1+\nabla\Phi_2+\nabla\Phi_\mathrm{p},
    \label{eq:motion}
\end{equation}

where $\Phi_1$, $\Phi_2$, and $\Phi_p$ are the gravitational potentials of the stars and the planet, respectively. The orbits of
the binary star and the planet are computed in terms
of the restricted three-body problem (i.e., we assume
that the planet does not affect the dynamics of the
binary star).

A symplectic algorithm \citep{1967PhRv..159...98V} was used to
integrate Eqs.~\ref{eq:motion}. The accuracy of this method is
proportional to $O(\Delta t^4)$, where $\Delta t$ is the time step.
The choice of a sufficiently small step $\Delta t$ allowed the
disk dynamics to be simulated on a time interval of $5 \times 10^4$ yr. To check the accuracy of our calculations,we performed selective integration of the constructed orbits backward in time, which showed good agreement with the initial data.

\section{Coorbital Dynamics}
\begin{figure}
    \includegraphics[width=\columnwidth]{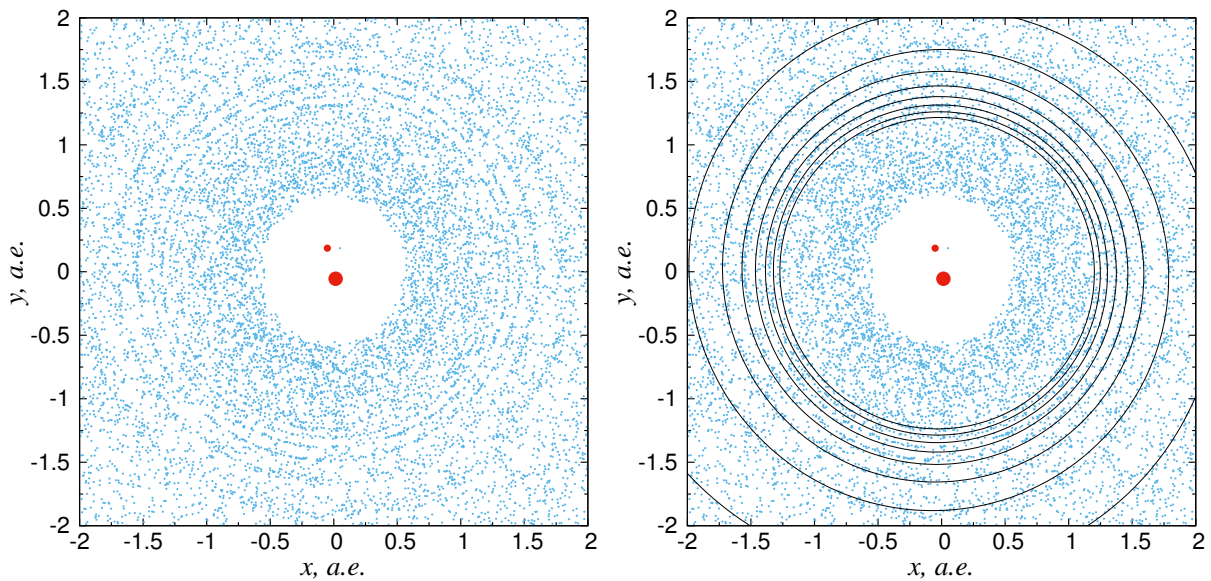}
    \caption{Distribution of particles in model 1 at $t=2\cdot10^3$~yr (a) and with a plotted analytical spiral (b). Here and below all distances are given in astronomical units.}
    \label{fig:K16}
\end{figure}

\citet{2015ApJ...805...38D} showed that
a one-armed spiral density wave could be formed in
the circumbinary planetesimal disk of a close binary
star. Our simulations of the motion of planetesimals
around a binary with the parameters of Kepler-
16 in model 1 also demonstrate the formation of
a spiral structure described by Eq. (6) from \citet{2015ApJ...805...38D} (Fig.~\ref{fig:K16}). The propagation
time of the spiral density wave for Kepler-
16 over a disk with a radius of $30$ AU is $T_s = 1.1 \times
10^7$ yr. The formation of a matter-free central cavity
is clearly seen in Fig.~\ref{fig:K16}. Its size is consistent with the numerical-experiment dependences and the theory \citep{1986A&A...167..379D, 1999AJ....117..621H, 2008IAUS..246..209V, 2015ApJ...799....8S}.

\begin{figure}
    \includegraphics[width=\columnwidth]{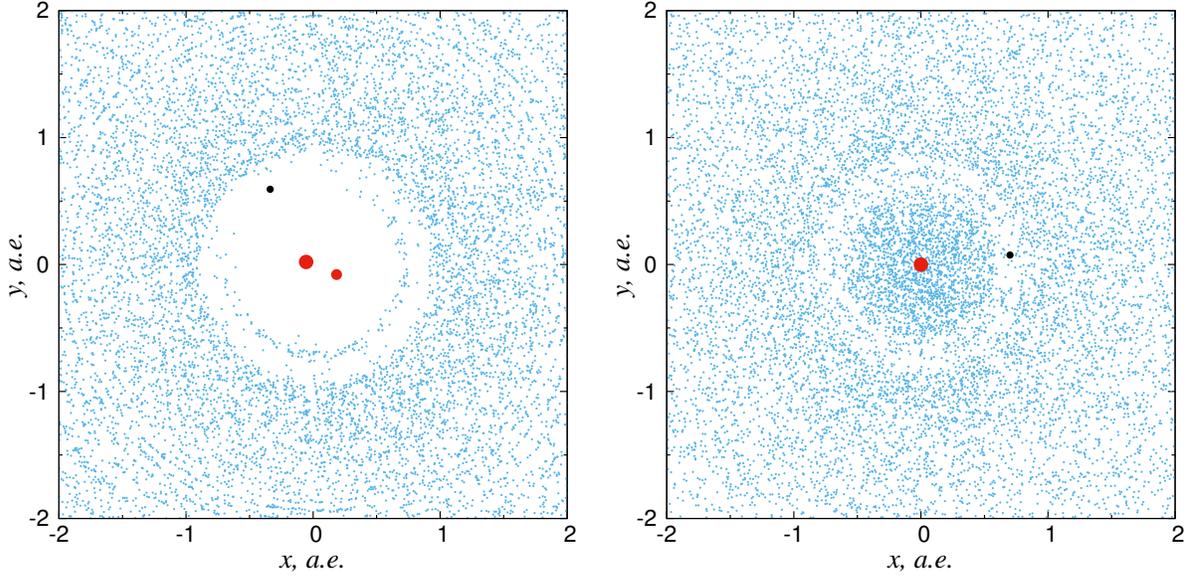}
    \caption{Distribution of particles at $t=10^4$~yr in models $2$ (a) and $3$ (b).}
    \label{fig:planet}
\end{figure}

The introduction of a planet that could be formed
by the time of gas disk depletion into the model destroys
the spiral structure near the planet's orbit, but
the spiral structure is retained on the extended-disk
periphery. The formation of a dense ring coorbital with
the planet from the planetesimal-disk matter can be
seen in Fig.\ref{fig:planet}. Recall that a similar ring also emerges in the models with a single central star \citep{2000ApJ...537L.147O}. The capture of particles into librational horse-shoe orbits near the Lagrangian points $L4$ and $L5$ \citep[on horse-shoe orbits, see, e.g.,][]{1999ssd..book.....M} is responsible for the formation of the
structure coorbital with the planet.

\begin{figure}
    \includegraphics[width=\columnwidth]{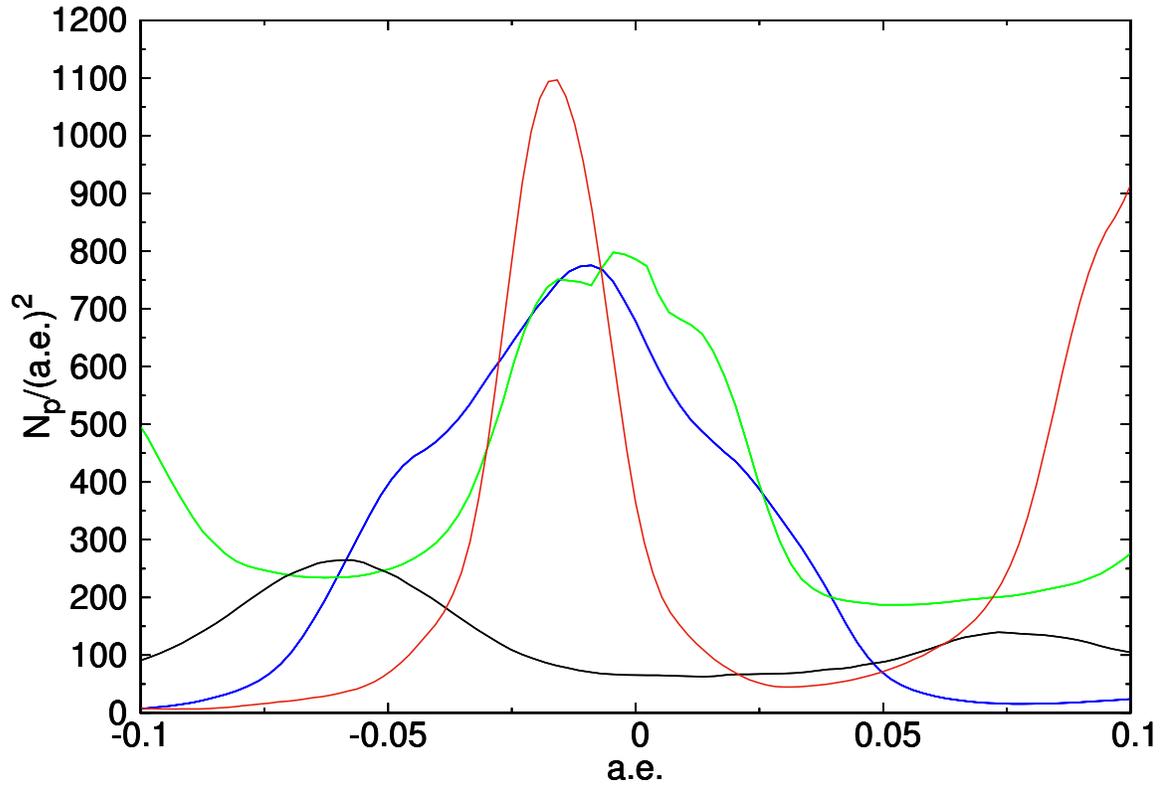}
    \caption{Azimuthally averaged radial barycentric surface density profiles in units [number of planetesimals$/$AU$^{-2}$] measured
from the radial position of the planet. The blue, green, black, and red lines correspond to models $2$, $3$, $4$, and $5$, respectively; $N_p/$(AU)$^2$ is the number of planetesimals per astronomical unit squared.}
    \label{fig:ring}
\end{figure}

To estimate the characteristics of the ring coorbital
with the planet, we constructed the azimuthally
averaged radial surface density profiles (Fig.~\ref{fig:ring}). The
planetesimal disk is binned along the radius into rings
with a step of $0.002$ AU. The number of planetesimals
within the rings was counted and then divided by the
ring area. In addition, these profiles were averaged
in time over a period from $10^4$ to $5 \times 10^4$ yr. Our
simulations showed that the coorbital ring structure
changes little in this time interval. The distances
from the radial position of the planet at which the
surface density changes by 1$\%$ of that at the center
of the coorbital ring were taken as the outer and
inner boundaries of the coorbital structure. Analysis
of the results showed that this approach allows the
attainment of a plateau by the surface density profile
to be revealed. This makes it possible to determine
the radial extent of the coorbital ring with confidence.

Within the revealed boundaries of the coorbital
ring the number of particles was calculated with a
time step of $10$ yr. In agreement with the results from~\citet{2016MNRAS.463L..22D}, our simulations
show that the number of planetesimals in the coorbital
ring in the single star system decreases more
rapidly than it does in the binary system. $95.9\%$ of the matter is retained in the coorbital ring for a time from $10^4$ to $5 \times 10^4$ yr in model $2$, while $71.5\%$ is retained in model $3$. In models $4$ and $5$, $77.9$ and $87.0\%$,
respectively, is retained.

The enhanced stability in the circumbinary case
probably stems from the fact that the presence of
a second star in the system causes a relatively fast
orbital precession of the planet and planetesimals. As a result, the splitting of mean-motion resonances
into subresonances is significant, they essentially do
not overlap, and the chaoticity of the motion near
the resonances is reduced sharply \citep[for examples of the action of such a mechanism, see][]{2014CeMDA.118..235E}.

The coorbital ring-like structure in model $2$ (the
blue line in Fig.~\ref{fig:ring}; model $2$ corresponds to Kepler-16) has a greater extent along the disk radius than does
the analogous structure in model $3$ (the green line in
Fig.~\ref{fig:ring}). The amount of matter within the boundaries
of the coorbital ring at the end of our simulations is
$1.19$ and $0.93\%$ of the total amount of matter in the
disk in models $2$ and $3$, respectively. However, $\sim 50\%$
of the particles in the ring-like structure in model $2$
are concentrated near the planet, while in model $3$ this
does not occur. A similar effect also takes place in
model $4$ but is absent in model $5$, which is probably
attributable to the small mass of the planet in model $5$.

In model $4$ (with the parameters of the binary
Kepler-34; the black line in Fig.~\ref{fig:ring}) the surface
density maximum is shifted toward the barycenter of
the binary orbit. In addition, the amount of matter
retained at the end of our simulations in the coorbital
structure under consideration is small, being $0.49\%$
of its total amount in the disk, while the surface
density is lower than that in other models by a factor
of $2-3$. This system is peculiar in that the initial
orbit of the planet is not close to a circular one, as
in the remaining systems under consideration; its
initial eccentricity is $e_p = 0.182$. The central binary
also has a significant eccentricity, $e_b = 0.521$. In all
likelihood, these two facts prevent the formation of a
stable coorbital structure.

The coorbital structure in model $5$ (Kepler-35)
is noticeably narrower than that in the remaining
models, with the maximum surface density being
higher than that in models $2$ and $3$ by a factor of 1.5.
The maximum of the radial distribution of matter is
slightly shifted relative to the radial position of the
planet. The compactness of the ring-like structure
is apparently attributable here to the small ratio of
the planet's mass to the sum of the binary component
masses $\mu_p = m_p/(M_1 +M_2) = 0.71 \times 10^{-4}$
(while for Kepler-16b $\mu_p = 3.34 \times 10^{-4}$). In addition,
in all likelihood, an approximate equality of the binary
mass components also stimulates the radial compactness
of the ring and an enhanced surface density
in it. The amount of matter contained in the coorbital
ring-like structure is $0.53\%$ of the total amount of
matter in the disk.

\section{Conclusion}
Our study of the dynamics of planetesimals in
the debris disks of the circumbinary systems Kepler-
16, Kepler-34, and Kepler-35 has shown that the
characteristics of the structures coorbital with the
planets in the disk depend significantly on the orbital
parameters of the binary system and the planet. For Kepler-34 as an example, it can be seen that a significant eccentricity of the binary star ($e_b \sim 0.5$) in combination with an appreciably noncircular orbit of the planet ($e_p \sim 0.2$) does not allow a stable long-lived coorbital ring to be formed. In contrast,
low or moderate eccentricities of the binary and the
planet facilitate the formation of a stable coorbital
structure. In all likelihood, the enhanced stabilization
of the coorbital ring in Kepler-35 is attributable
to an approximate equality of the binary component
masses. Our simulations show that the formation
of a noticeable coorbital structure begins after $\sim 500$
orbital revolutions of the planet (which corresponds
to a time interval of $\sim 300$ yr for Kepler-16b).

Thus, the observational detection of planetesimal
rings coorbital with the planets in Kepler-16 and
Kepler-35 as well as in other systems with similar orbital
parameters is quite probable. The observational
detection of such ring-like structures in circumbinary
disks without resolved planets would suggest the real
existence of planets generating these structures. The
spiral density wave described theoretically by~\citet{2015ApJ...805...38D} is destroyed near the
planet's orbit but can be noticeable on the disk periphery.

The stability of the ring-like coorbital structures in
circumbinary systems is much higher than that of the
coorbital rings for planets of single stars. Therefore,
the observational detection probability of such structures
is also higher in the case of binary star systems.\\

\textbf{Acknowledgments}. We are deeply grateful to the referees for their useful remarks. This work was supported by the Russian
Foundation for Basic Research (project no. 17-02-
00028-a) and the ``Solar System'' Program no. 7A of
the Presidium of the Russian Academy of Sciences.

\bibliography{biblio}
\end{document}